# How do Range Names Hinder Novice Debugging Performance?

Ruth McKeever, Kevin McDaid

Dundalk Institute of Technology,

Dundalk, Ireland.

catherineruth.mckeever@dkit.ie, kevin.mcdaid@dkit.ie

**ABSTRACT**

*Although experts diverge on how best to improve spreadsheet quality, it is generally agreed that more time needs to be spent testing spreadsheets. Ideally, experienced and trained spreadsheet engineers would carry this out, but quite often this is neither practical nor possible. Many spreadsheets are a legacy, developed by staff that have since moved on, or indeed modified by many staff no longer employed by the organisation. When such spreadsheets fall into the hands of inexperienced, non-experts, any features that reduce error visibility may become a risk. Range names are one such feature, and this paper, building on previous research, investigates in a more structured and controlled manner the effect they have on the debugging performance of novice spreadsheet users.*

**1 INTRODUCTION**

The collapse of financial systems worldwide has brought the lack of regulation to the forefront of public anxiety. In many cases spreadsheet errors have caused significant financial losses for organisations, leading to market uncertainty. One example of this is the case of C&C (EuSpRIG 2010) where shares fell 15% "after data were incorrectly transferred from an accounting system used for internal guidance to a spreadsheet used to produce the trading statement." Such occurrences establish the need for protocols and best practices to reduce these errors, and for increased awareness of how often errors occur. Panko (2003) notes, "In spreadsheeting, developers who do not do comprehensive error checking are rewarded both by finishing faster and by avoiding onerous testing work."

At EuSpRIG'09 a paper (McKeever, McDaid et al. 2009) was presented that challenged the commonly held view that range names improve the quality and understandability of a spreadsheet. The motivation for this study was a lack of empirical evidence to support these views. The authors questioned several aspects of range names, including how they are viewed by academia and industry, if they are used in practice, and how they affect the debugging performance of novice spreadsheet users. This publication created considerable debate on the risks associated with prescribing techniques that had not been tried and tested. It also raised concerns about the appropriate use of names, and whether the trial accurately reflected how names are used in practice.

The findings of this exploratory study show a decrease in the ability of novices to identify and correct formula errors in a spreadsheet when range names are used extensively. The feedback indicates that the level of use of names may have been too high, and the new study seeks to establish if the problems identified with range name usage continue to occur when the volume of names in the spreadsheet are kept to a minimum. Based on the results and feedback from the initial study the new trial was modified to tackle the limitations and issues raised. It was then carried out on two groups of computing students in order to conduct a more comprehensive study of novice spreadsheet users. The results of this controlled experiment are in keeping with the findings of the exploratory experiment, and go further towards dispelling the notion that named ranges make spreadsheets easier for novice users.

In a spreadsheet, a range is defined as a cell or group of cells. A range name is a name given to a range, which can then be used throughout the workbook in place of the cell references. In addition to cells, the developer can also name constants, and use this name as they would use a variable in software code. The main advantage of this for the developer is that they can change what the name refers to in one place instead of trawling through all the formulas in which it is referred to. In addition, it is believed that by replacing a cell reference with a meaningful name it will make the formula more understandable to the user.

Names are a powerful feature of Excel, and used properly can prove very useful to developers. For example, Grossman et al. (2009) show how range names can be used to replace complex nested-if formulas with the lookup technique. As nested-if formulas are considered risky, the lookup technique uses names to make the logic required simpler, more visible and therefore less risky.

To name a range the user highlights the range then enters a name in the name box (the box above and to the left of the worksheet that normally contains the reference of the cell that is currently in focus). To modify or delete this name, or to create a named constant, the developer must use the name manager, contained in the formula section of the Excel 2007 ribbon. This also provides the facility to insert a descriptive comment for the name. To access a named range in a workbook the user may either choose from the list of names in the name box, or use the F5 key to bring up a list of names contained in the workbook. This does not work for constants, however, as they are not linked to cells in the workbook and must be accessed through the name manager.

Several commercial tools exist that provide extensive support for range names. One such tool is OAK (OPERIS Analysis Kit) (OPERIS 2009): "OAK can get rid of the names, leaving a coordinate-only version of the model to be checked". Spreadsheet Detective (Berglas and Berglas) also contains several useful functions for working with range names, including *annotate* which displays a box above each formula cell showing what each name refers to, and *precedent/dependent dialog* which allows the user to navigate through cell precedents and dependents.

Range names are advocated by many in both academia and industry, including the SSRB (Spreadsheet Standards Review Board) (BPM Analytical Empowerment 2005), and Microsoft (Microsoft 2006), (Microsoft 2008). Examples of such advice can be found as far back as 1986 (Bromley 1986). Bewig (2005) advocates the proper construction of range names for eliminating the problem of referring to the wrong cell while constructing formulas, and states that "well-chosen names are the first and best form of documentation." Campbell (2009) recommends "To allow for easily-adapted spreadsheets, make sure nothing is hard-coded in terms of locations – any cells being used in formulas should be referred to as named ranges." Kruck (2006) describes a formula using names that "will be easier to understand and maintain for future developers". Many of the popular spreadsheet advice websites also advise users on how to use range names, such as (Pearson 2009), (Ozgrid 2008) and (MrExcel 2007).

Not all experts are convinced about the merits of range names. Panko and Ordway (2005) recognise the risk that range names may appear correct, while actually referring to the wrong range, warning that "although the research findings are not clear on this issue, using range names should be considered potentially dangerous until research on using range names is done". Blood (2006) also lists several additional problems with range names, such as increasing the length of formulas, and creating "ghost" links when a sheet is copied to another workbook. Butler (2000) describes a risk assessment methodology developed by HM Customs & Excise in response to the material quantitative errors found in 10% of spreadsheets tested that belong to taxpayers. This methodology identifies range names as a high-risk feature, along with macros and hidden rows/columns, when assessing the risk of the application.

## 1.1 Overview

Section 2 describes the exploratory experiment that this study extends. Section 3 details the methodology, rollout and results of the experiment central to this new work. Section 4 discusses the results and limitations of this study. Section 5 details future work arising from this study, and Section 6 concludes the paper.

## 2 SUMMARY OF EXPLORATORY STUDY

The studies described here are based on an experiment originally designed by Howe and Simkin (2006), and used by Bishop and McDaid (2007). It involves asking a group of students to examine a spreadsheet and to correct any errors that they find. The main advantage of using a well-established experiment for testing a new theory is that there is existing data with which to compare the findings.

The exploratory study as detailed by McKeever et al. (2009) analyses how a group of 21 students in their second year of a computing degree in Dundalk Institute of Technology perform debugging a spreadsheet seeded with errors that was developed to include named ranges. The results of the experiment are then compared with an earlier experiment (Bishop and McDaid 2007), developed without named ranges, carried out on 34 second year accounting and finance students. These experiments are identical in format, except for the inclusion of named ranges in the 2009 trial. The spreadsheet contains 42 seeded errors, categorised as clerical/non-material (4), rule violation (4), data entry (8) and formula (26). The formula errors are further divided into logic (9), cell reference (7), range reference (7) and remote reference (3). A recording tool is used in these trials to track the cells entered by the user, the time spent in each cell, and any changes made to the spreadsheet.

## 2.1 Results

The results of the exploratory study revealed that both groups had performed almost identically when correcting the first three categories of errors, but the group that used named ranges corrected significantly fewer formula errors. This was true for all sub-categories of formula errors, but was most pronounced for *cell* and *range reference* errors. These results are shown in Table 1.

> While the experiment group performed worse in all the formula sub-categories, the greatest difference was with *Range* and *Cell reference* errors, and these were the only groups where the results were statistically significant at a 5% level based on a non-parametric rank sum test. In each case of these types of error the control group performed better than the experiment group. (McKeever, McDaid et al. 2009)

| Error Type | No of Seeded Errors | % Corrected by Named Group | % Corrected by Unnamed Group | Named Compared to Unnamed |
|---|---|---|---|---|
| Clerical/Non-Material | 4 | 11% | 11% | 0% |
| Rule Violation | 4 | 63% | 65% | -2% |
| Data Entry | 8 | 64% | 63% | 1% |
| Formula | 26 | <u>44%</u> | <u>63%</u> | <u>-19%</u> |
| *Logic* | *9* | *54%* | *63%* | *-9%* |
| *Cell* | *7* | <u>*39%*</u> | <u>*68%*</u> | <u>*-29%*</u> |
| *Range* | *7* | <u>*47%*</u> | <u>*71%*</u> | <u>*-24%*</u> |
| *Remote* | *3* | *19%* | *28%* | *-9%* |

**Table 1 – Exploratory Study Results**

## 2.2 Discussion

Analysis of the possible explanations for the differences in performance focussed on the following three possible causes.

*High cognitive load*: Cognitive overload occurs when a person is overwhelmed with information, and cannot take it all in. Working with spreadsheets requires a high level of working-memory cognitive skills, such as memory load (Kruck, Maher et al. 2003). A range name is an additional piece of information that the user must remember. As the participants were unlikely to remember what each name referred to, they would have to perform two checks, one to see if the correct name was used and another to see if the name referred to the correct range. This theory is difficult to measure, but its impact can be assessed by examining the correction rates of errors caused by incorrect use of names and errors in cells that contain names, with formula errors in cells that do not contain names. Kruck et. al. (2003) found that spreadsheet training improves logical reasoning cognitive skills, and this decreases error rates. Tukiainen (2001) evaluated spreadsheet calculation under the cognitive dimensions questionnaire and found that "when asked what kind of things are more difficult to see or find, most of the subjects mention formulae or referencing of the cells in formulae." Range names add another layer of concealment to already hidden formulae.

*Too much confidence in names*: There was some evidence that if there was more than one name used in a formula then the participant, once satisfied that the first name was correct didn't bother to check subsequent names. Overconfidence is recognised as a serious issue in spreadsheet development (Panko 2003), and this finding indicates that range names increase the trend.

*Did not understand the error/did not know how to correct it*: Although this was investigated as a possible cause, this was not supported by the results. If a participant made one or more attempts to fix an error they tended to succeed, indicating that they knew how to correct the errors identified. The participants had also been observed completing a tutorial on named ranges before the trial, to ensure they could complete the task.

**2.3 Limitations**

While for the purposes of an exploratory study the use of a different experiment group for comparison was deemed sufficient, the use of a control group would considerably improve the legitimacy of these results.

The spreadsheet contains 152 range names. This is considered excessive and not realistic, as is the length of many of the names (e.g. VariableExpensesTotalYearEstimate). This may increase the cognitive load on the participants. Another problem identified with this extensive use of names is that in some cases it becomes impossible to copy formulas down a column. This reduces the convenience of spreadsheets as a modelling tool.

Before commencing the trial, the participants were given a tutorial on how to name ranges, and how to edit those names. This may have influenced their opinion of names and impressed upon them the advantages of naming. This could explain the finding of overconfidence.

As these limitations may exaggerate the causes identified, they are addressed in the new version of this trial, as described in this paper.

**2.4 Hypotheses**

A new study was planned to reassess the findings of the exploratory study, overcome the limitations, and investigate a further set of more detailed hypotheses. The overall hypothesis is as follows:

*Novice debuggers perform less well in correcting formula errors in a spreadsheet that contains range names.*

This will be investigated by exploring the types of range name and non-range name errors in formulas through the following more detailed hypotheses:

*H1: Novice debuggers perform less well in correction of cell formulas where the error is due to the wrong range being assigned to a name, than if a name is not used in the formula.*

*H2: Novice debuggers perform less well in correction of cell formulas where the error is due the wrong range name being used in a formula, than if a name is not used in the formula.*

*H3: Novice debuggers perform less well in correction of cell formulas where the formula contains a name, but the error is not due to the name.*

*H4: Novice debuggers perform less well in correction of cell formulas, where there are no names in the formula, but names in the spreadsheet.*

Note that the above represent four distinct hypotheses and, in the context of the experiment described and associated statistical analysis, these represent the form of the alternative hypotheses with the null hypotheses stating that there is no difference in performance.

## 3 EXPERIMENT

The exploratory experiment allows the authors to frame a controlled study around a more detailed set of hypotheses and to overcome the limitations as discussed. This controlled study is kept largely the same as the original, but with the number of names reduced, and was performed on Computing students in Dundalk Institute of Technology.

### 3.1 Methodology

**Hypotheses**

Errors due to range names can be divided into two categories: first are the errors that occur when the wrong data or range is assigned to a name; second are the errors that occur when the wrong name is used in a formula. In the first case it is clear that names actively hide the error. This increases the cognitive load on the user and may lead to a lower discovery rate. The second case, where the wrong name is used, may well result in a better performance by debuggers – as the error may be more visible due to the fact that the main advantage of names is that that they make explicit to the user what is happening in the formula. If user fails to spot this type of error it may indicate overconfidence. In addition to these categories of errors, this trial also contains both formula errors that are themselves not related to names, but are in a cell or formula that contains one or more names, and formula errors that occur in a cell that does not contain range names.

Hypotheses 1 through 4 will be addressed by comparing the correction rates for relevant errors, between the experiment and control groups.

*H1: Novice debuggers perform less well in correction of cell formulas where the error is due to the wrong range being assigned to a name, than if a name is not used in the formula.* This occurs in four errors, numbered 27, 28, 29 and 37. If this is confirmed it will support the theory that cognitive load is increased by the use of names and that names have a negative impact on the debugging performance of novice spreadsheet users.

*H2: Novice debuggers perform less well in correction of cell formulas where the error is due the wrong range name being used in a formula, than if a name is not used in the formula.* This occurs in two errors, 18 and 35. If this is confirmed it will suggest that overconfidence is an issue, as these are the errors that names are supposed to make most obvious.

*H3: Novice debuggers perform less well in correction of cell formulas where the formula contains a name, but the error is not due to the name.* Three formula errors occur under these circumstances, 13, 33 and 36. If this is proven it will indicate that names distract the user from other possible errors.

*H4: Novice debuggers perform less well in correction of cell formulas, where there are no names in the formula, but names in the spreadsheet.* There are 13 errors in the trial spreadsheet that are neither caused by, nor in the same cell as, a name. If this hypothesis is proven it will suggest that either the inclusion of names distracts the user from debugging the entire spreadsheet, or that the inclusion of names makes the user over-confident in the whole spreadsheet.

**Design**

The following changes were made to the design of the exploratory experiment, in order to address the limitations identified:

- A control group was used to improve the validity of this experiment.
- The number and length of names in the experimental workbook were reduced.
- The students were not given a tutorial on naming ranges.

The number of errors was reduced from 42 to 39. One error was removed each from the control group, and the experiment group, as these errors were not comparable between the groups. This brings the total error count down to 38, as categorised in Table 2. This reduction in errors is partly due to a reassessment of what constitutes an error. Previously each cell that contained an error was considered to be an error. Each error is now considered independently, as a single error can occur in more than one cell, and a cell can contain more than one error. The results for this study are based on the 38 individual errors.

The number of names was reduced from 152 to 12 - 10 named cells and 2 named constants. This reduces the cognitive load, as there are fewer names to consider. On reviewing the literature on naming in software development it was decided to use full words when naming ranges, carefully chosen to provide information to the user. Rowe (1985) describes good names as "evocative, and which conjure up distinctive characteristics of what is being named." Abbreviations are not used, as they would require domain knowledge for the user. Any names that could prevent the user from dragging a formula down a column or across a row were also removed, and arrays were not used. It is the belief of the authors that this spreadsheet is a more accurate reflection of how a developer might use names in practice.

The students were not given a tutorial on range names as their lecturer confirmed that they had been taught how to use them, although they were reminded how to access the name manager. The advantages of names were not impressed on them at the time of the trial, in order to reduce overconfidence.

**Sampling**

The redesigned experiment was conducted on two groups of computing students in a Dundalk Institute of Technology computer laboratory. The first group was made up of 14 students; the second group, 15. To ensure the validity of the research, each participant was assigned randomly to a control or experiment group.

**Materials**

Each student was given an instructions sheet detailing the rules and assumptions that the data in the spreadsheet was expected to follow, a consent form following the DkIT ethics guidelines, and the spreadsheet model for the experiment. The spreadsheet model contained T-CAT, a "time-stamped cell activity tracking tool", (Bishop and McDaid 2008) to monitor their cell clicks, changes and times.

**Process**

The students were given a brief introduction to the research, talked through enabling macros in Excel settings, and shown where to find the name manager. The instructions and consent sheet were distributed to the participants, and the workbook required for the experiment was saved to each of their PCs, according to whether they were part of the control or experiment group. Each participant renamed the workbook to their own name, and when they had completed the task the workbooks were collected, along with signed consent forms.

**Method of Analysis**

The resulting workbooks were processed through a macro that looked at the values in all error cells, to deduce if the participant had correctly identified and corrected the error. The values in the cells that were returned as incorrect were manually examined to establish if the participant had used an unexpected way to correct the error. From the first group of students there were no results for one student, assigned to the experiment group, as he had returned an incorrect workbook. Another student from this group, assigned to the control group, failed to interact properly with the task, and found only one error despite making many changes to the spreadsheet; this data was removed from the overall results.

**3.2 Results**

Overall the control group corrected 66% of errors and the experiment group corrected 59% of errors, i.e. the experiment group found 6% fewer errors than the control group. When this is divided into error categories, the experiment group performed consistently worse at finding formula errors. This data is shown in Table 2.

| **Error Type** | **No. Of Errors** | **Experiment** | **Control** | **Difference** | **P-Value (1 Sided Rank Sum Test)** |
|---|---|---|---|---|---|
| Clerical | 4 | 23% | 28% | -4% | 0.26 |
| Rule Violation | 4 | 71% | 60% | 11% | 0.81 |
| Data Entry | 8 | 63% | 66% | -4% | 0.29 |
| Formula - Logic | 6 | 73% | 82% | -9% | 0.14 |
| Formula - Reference | 7 | 69% | 84% | -15% | 0.12 |
| Formula - Name | 9 | 51% | 60% | -9% | 0.16 |
| **Average** | 38 (Total) | **59%** | **66%** | **-6%** | **0.05** |

It is interesting to note that participants in the experiment group performed worse on all formula errors (logic errors -9%, and cell, range and remote reference errors -15%), not just errors involving names. This is in keeping with the findings of the exploratory study. The results of the errors that specifically relate to names, and are directly comparable between the control and experiment groups, also reveal that the experiment group found 9% fewer errors (control group: 60%, experiment group: 51%). Name errors are defined as those where an incorrect range is assigned to a name, a name is used incorrectly in a formula, or an error occurs in a formula that also uses a name.

H1 is dependent on the correction rates of errors, 27, 28, 29 and 37, as shown in Table 3. While the experiment group performed consistently worse than the control group in these errors, they did not perform as badly as expected, based on the results of the exploratory study. It appears that the reduction of names in this trial had a positive effect in this respect. Although the difference in correction rates for each error is not statistically significant (p-value 0.46), it provides weak support for the hypothesis. Any support for this hypothesis would say that high cognitive load impacts on the experiment groups' performance, as these are the errors that require a double check, one to ensure the correct name is used, and another to see if the name refers to the correct range.

| Wrong range assigned to name | | | | |
|---|---|---|---|---|
| **Error No.** | **Location** | **Experiment** | **Control** | **Difference** |
| 27 | Office Expenses F10 | 43% | 50% | -7% |
| 28 | Office Expenses F18 | 79% | 80% | -1% |
| 29 | Office Expenses F20 | 36% | 40% | -4% |
| 37 | Projections B9/B10 | 36% | 40% | -4% |
| **Average** | | **48%** | **53%** | **-4%** |

Table 3 - Hypothesis 1 Errors

H2 depends on the correction rates of errors 18 and 35, as shown in Table 4. The experiment group performed 34% worse than the control group on Error 18. This error occurred when the formula "=Subtotal_A+Subtotal_B+Subtotal_B" was entered instead of "=Subtotal_A+Subtotal_B+Subtotal_C". This repeats a finding from the exploratory experiment that if there is more than one name in a formula the participant will not continue to search for errors beyond the first name. In stark contrast to this result, the experiment group performed 27% better at identifying error 35. This error occurred when two names were entered in the wrong order in cells one above the other. While this looks like a positive result, every participant of the experiment group failed to identify the second error in the same cells (Error 36, shown in Table 5). This is in keeping with the findings of the exploratory experiment, where no participants corrected both errors. This appears to be linked to over-confidence, as once the experiment group had identified one error they stopped looking for others, although this is not peculiar to spreadsheets containing names.

While the results of these errors provide very limited support for Hypothesis H2 (again not statistically significant), they might suggest that when a formula contains a name error the user can be distracted from identifying other errors in the same cell. Nonetheless this result does not conclude, as one might expect, that range names would help users to debug formulas. More studies need to be carried out to support this finding, using more examples of these types of errors.

| Wrong name used in formula | | | | |
|---|---|---|---|---|
| Error No. | Location | Experiment | Control | Difference |
| 18 | Payroll I6 | 36% | 70% | -34% |
| 35 | Projections B19/B20 | 57% | 30% | 27% |
| **Average** | | **46%** | **50%** | **-4%** |

Table 4 - Hypothesis 2 Errors

H3 is illustrated by three errors: 13, 33 and 36. These errors are unrelated to names, but which occur in cells that also contain names, and are shown in Table 5. While all these errors show that the experiment group performed worse, errors 13 and 33 have a high correction rate to begin with. The low correction rate for error 35 can be in part explained by the occurrence of two errors in the cell, as shown in Table 4. The difference in performance is statistically significant (p-value = 0.04) and thus the results support both the hypothesis and the theory that names increase over-confidence, as the participants were less likely to look for other errors if a name was contained in the cell.

| Other formula errors in cells that also contain names | | | | |
|---|---|---|---|---|
| Error No. | Location | Experiment | Control | Difference |
| 13 | Payroll G11 | 86% | 90% | -4% |
| 33 | Projections B17 | 86% | 100% | -14% |
| 36 | Projections B19/B20 | 0% | 40% | -40% |
| **Average** | | **57%** | **77%** | **-20%** |

Table 5 - Hypothesis 3 Errors

H4 depends on all the other formula errors in the spreadsheet. These are shown in Table 6. In ten out of thirteen cases the experiment group performed worse than the control group and the difference in performance is statistically significant (p-value = 0.04). Note that the p-values throughout this paper are based on a one-sided rank sum test as it is judged that an assumption of normality may be unrealistic for this discrete data. The result provides strong support for Hypothesis H4.

| Other formula errors in cells that do not contain names | | | | |
|---|---|---|---|---|
| Error No. | Location | Experiment | Control | Difference |
| 8 | Payroll F9 | 57% | 90% | -33% |
| 9 | Payroll F10 | 93% | 90% | 3% |
| 10 | Payroll F11 | 57% | 80% | -23% |
| 11 | Payroll G6 | 57% | 80% | -23% |
| 12 | Payroll G7 | 43% | 80% | -37% |
| 14 | Payroll G16 | 79% | 90% | -11% |
| 15 | Payroll H16 | 79% | 90% | -11% |
| 16 | Payroll I10 | 79% | 80% | -1% |
| 17 | Payroll I14 | 71% | 80% | -9% |
| 25 | Office Expenses E8 | 64% | 90% | -26% |
| 26 | Office Expenses F5 | 86% | 80% | 6% |
| 38 | Projections G17 | 93% | 70% | 23% |
| 40 | Projections G22 | 64% | 80% | -16% |
| **Average** | | **71%** | **83%** | **-12%** |

Table 6 - Hypothesis 4 Errors

# 4 DISCUSSION

One of the main issues with the first exploratory experiment was the over-use of names in the workbook, and the length of those names. This may have contributed to the participants' confusion, and distracted them from finding the errors. Furthermore the selection of a different study for the control group was not ideal.

In this experiment we have addressed these issues. We have also tested hypotheses regarding debugging performance for four scenarios regarding names and errors in formulas.

Despite significantly reducing the number of names in the workbook, and the length of the names, the experiment group still performed worse than the control group at identifying and correcting formula errors. Furthermore the group performed worse in the correction of errors in formulas including names and formulas not including names. An obvious statement may be that the participants performed poorly across all categories of formula errors because of a lower ability. However, participants were randomly assigned to groups and, importantly, their performance in the other categories of *clerical*, *rule violation* and *data entry* error was broadly similar, indicating equal ability.

Based on the first experiment it was considered that the poor performance of the experiment group may be due to three possible factors: high cognitive load, over confidence and insufficient knowledge. Insufficient knowledge was dismissed as a factor, after extensive analysis of the results, including analysis of time and attempts made to correct errors.

Through the hypotheses presented, this experiment provides some support for the first two possible causes: high cognitive load, and over-confidence. In particular, the finding that the inclusion of names seems to distract the user supports the high cognitive load theory. Names hide errors, and are simply another detail that has to be remembered by the user.

The results of the errors show that range names do not improve debugging performance, and go some way towards supporting the hypotheses. This indicates that the greatest problems with names lie not in their implementation, but in the effect their inclusion has on users' perception of the spreadsheet.

## 4.1 Limitations

This study was conducted on novice spreadsheet users, with a background in computing, and can therefore only be applied to such users, although previous experiments have shown that the performance of computing students is very similar to that of accounting and finance students (Bishop and McDaid 2007; Bishop and McDaid 2008).

There are two errors in this experiment that cannot be compared between the two groups; one in the experiment workbook (Error 30), the other in the control (Error 34) workbook. This was an oversight and will be rectified in any future iterations of this experiment.

One problem identified with the name errors seeded in this trial is that these errors have a particularly low correction rate across all trials. Only 56% of the participants in (Bishop and McDaid 2007) corrected these errors, compared with 69% who corrected the remaining formula errors. Likewise, only 37% of the participants in (McKeever, McDaid et al. 2009) corrected these errors, while 52% corrected the remaining formula errors.

However, this study examines difference in performance between groups rather than actual performance of groups.

**5 FUTURE WORK**

The big question that has yet to be addressed is how professionals perform when debugging spreadsheets that use named ranges. One of the main purposes of this experiment was to improve the design to a stage where it is worthwhile rolling out to professionals. The following hypothesis will be addressed in the next phase of this research:

*Professional debuggers perform less well in correcting formula errors in a spreadsheet that contains range names.*

This experiment will be distributed to a group of experts, possibly by email, and the same analysis performed. The results will be compared with those from the novice groups. A control group is vital here as it is to be expected that experts will find more errors than novices overall. This hypothesis will be confirmed if the difference in correction rates between the professional control and experiment groups is comparable to the difference in correction rates between the novice control and experiment groups.

Another aspect that has not been looked at is how the participants rate their debugging skills. Any future experiments will include questionnaires to measure how confident the participants are in the accuracy of their debugged spreadsheet.

An experiment to investigate whether names have an impact on error rates during development is planned. This will establish if spreadsheet engineers should be advised to develop using names, but remove them before the spreadsheet is used.

**6 CONCLUSION**

Building on the paper presented at EuSpRIG'09 (McKeever, McDaid et al. 2009), this work describes a well structured and controlled experiment that provides additional support for the theory that named ranges make it more difficult for novices to debug spreadsheets. The structure is provided through four hypotheses linked to types of formula errors.

The study provides strong (statistically significant) evidence to support the hypotheses (H3 & H4) that the inclusion of range names in a spreadsheet impacts negatively on the performance of novice debuggers even for formulas where the error is not due to an error in a range name.

For the hypotheses (H1 & H2) where the error is due to an error in a range name the evidence to support the claim is not as strong. Although the results were not statistically significant, it can be said that the experiment certainly does not provide support for the argument that range names help novices to debug formulas with range names. In this context, this contradicts the claim that range names make spreadsheets easier to understand.

The results of this trial are consistent with the results of the original exploratory study. Removing a large volume of names marginally improved the debugging rate, but not to the level of the control groups. This dispels the idea that it is just the misuse and overuse of names that cause novice debuggers to find fewer errors, although these issues clearly have an impact.

It is our belief that it is simply not sufficient to reduce names; they should be eliminated altogether unless absolutely necessary. Range names simply do not help novice debuggers. While over-confidence and high cognitive load may be a problem inherent to spreadsheets, they are clearly exacerbated by range names. There is technology that replaces range names with the base range and it is our belief that this should be used for debugging by novices.

Finally, our attention will now turn to experts and their performance and behaviour when debugging spreadsheets that contain range names.